\documentclass[conference]{IEEEtran}
\IEEEoverridecommandlockouts
\usepackage{cite}
\usepackage{amsmath,amssymb,amsfonts}
\usepackage{algorithmic}
\usepackage[ruled,vlined]{algorithm2e}
\usepackage{graphicx}
\usepackage{textcomp}
\usepackage{multirow}
\usepackage{caption}
\usepackage{subcaption}
\usepackage{makecell}
\usepackage{xcolor}
\usepackage{hyperref}
\usepackage{url}
\def\BibTeX{{\rm B\kern-.05em{\sc i\kern-.025em b}\kern-.08em
    T\kern-.1667em\lower.7ex\hbox{E}\kern-.125emX}}
\begin{document}

\title{Detecting Timing Attack on PMU Data utilizing Unwrapped Phase Angle and Low-Rank Henkel Matrix Properties\\
}

\author{\IEEEauthorblockN{Imtiaj Khan}
\IEEEauthorblockA{\textit{Electrical and Computer Engineering} \\
\textit{Virginia Tech}\\
Blacksburg, VA, USA \\
imtiajkhan@vt.edu}
\and
\IEEEauthorblockN{Virgilio Centeno}
\IEEEauthorblockA{\textit{Electrical and Computer Engineering} \\
\textit{Virginia Tech}\\
}

}

\maketitle

\begin{abstract}
Introduction of PMUs to cyber-physical system provides accurate data acquisition, while posing additional risk of being the victim of cyber attack. Both False Data Injection Attack (FDIA) and GPS-spoofing or timing attack can provide malicious data to the cyber system, though these two attacks require different post-attack contingency plan. Thus accurate detection of timing attack and separating it from conventional FDIA has become a very important research area. In this article, a successful detection of timing attack mechanism is proposed. Firstly, a method to distinguish timing attack and FDIA using unwrapped phase angle data is developed. Secondly, utilizing low rank Henkel matrix property to differentiate timing attack from electrical events is also presented. Finally, an experimental validation of proposed model is performed on IEEE 13 bus system using simulated GPS-spoofing attack. It can be observed that the timing attack can increase the rank 1 approximation error of Henkel matrix of unwrapped angles by 700 $\%$ for 3 sec delay in GPS time-stamp. The rank 1 approximation error is increased by 500$\%$ for 2 sec delay and the increase is insignificant for 1 sec delay attack. FDIA doesnt show any significant change in the low rank approximation profile of Henkel matrix.
\end{abstract}


\section{Introduction}

Inclusion of smart devices and integration of the physical power system with cyber system have put the whole power system at the risk of cyber attacks \cite{10}. Phasor Measurement Units (PMUs) are smart devices which measure signals at specific bus locations of the grid and provides time synchronized voltage and current phasor data to Phasor Data Concentrator (PDC) \cite{12}. The time synchronization with the Coordinated Universal Time reference (UTC) is done using a GPS 1 Pulse-Per-Second (PPS) and a time-stamp. Most PMUs provide data at rates between 30 and 120 sample per second with slower rates available \cite{13}. Due to its sophisticated nature, PMU data is prone to unique malicious attack by hackers \cite{11}.\\
The most common types of attack is the False Data Injection Attack (FDIA), where the attack deliberately inject falsified measurements into the actual PMU measurements. These types of attacks are aimed to force the control center into taking wrong decision based on the received measurements \cite{14}. Various researchers have proposed different FDIA detection mechanisms, the most common types is residual based detection method, which is not robust against coordinated stealthy FDIA \cite{15}. Other methods have been proposed to detect stealthy FDIA. In \cite{16}, a one cosine similarity metric based detection method is proposed to compute the difference between the estimated values and the measurements. An adaptive cumulative sum based method is proposed in \cite{17}, where the changes of residual vector mean are detected.\\
Generally due to the low level security of the civilian GPS signals, attackers superimpose falsified signal to the GPS in what is known as GPS-spoofing attacks. During GPS-spoofing, the attacker can manipulate GPS-clock and create false timestamp or time shift the 1 PPS disrupting the PMU time synchronization \cite{18}. This types of attacks can be referred as \textit{timing attack}. Though the detection and prevention of False Data Injection Attack are frequently covered in contemporary literatures, timing attacks have received little attention. In this work undetectable timing attack model has been developed where the attack can be bypassed the conventional FDIA detection method \cite{19}. The timing attack is modeled as an FDIA where only the phase angle data are manipulated by the attacker, since GPS spoofing shifts the time-reference and therefore changing the phase angle of voltage and current data. The authors in \cite{19} showed that at least two phase angle data from two separate PMUs need to be manipulated to create stealthy timing attack. The GPS spoofing can be easily detected if the defender is connected to another trusted GPS which contains encrypted military code \cite{18}. This method is not feasible for civilian infrastructure and large power grids. Therefore accurate detection of timing attack or GPS spoofing attack is necessary.\\
Timing attack can be detected using same method as FDIA detection since timing attack can be considered as an FDIA on phase angle data. However, this method fails to distinguish between these two types of attacks. As the timing attack needs different post-attack contingency plan than the FDIA, it is imperative for the cyber system to know the attack type. Moreover, electrical events such as line outage, oscillation event, frequency event, transformer faults can also cause incorrect PMU data. Thus, timing attack needs to be separated from electrical event too. Few research works have been carried out to separate event from attack, such a decision tree based method is proposed in \cite{20} to differentiate bad data from events. Utilizing low rank Henkel matrix property can also be a possible solution toward separating event data from cyber attack \cite{2}.\\
The aim of this paper is to successfully detect timing attack and distinguish it from FDIA. Raw phase angle data don't provide enough insight for this purpose since both types of attacks change the temporal relation between PMU channels. Instead of using raw data, unwrapped phase angle data can provide different temporal behavior for FDIA and for timing attack. In order to know if the difference in the unwrapped phase angle data is due to an attack and not due to the random time-series variation, the low rank approximation of Henkel matrix can be exploited. Therefore, the contributions of this paper are as follows:
\begin{itemize}
\item Using unwrapped phase angle data to differentiate between FDIA and timing attack.
\item Utilizing low-rank approximation of Henkel matrix to successfully detect the timing attack and to distinguish it from electrical events.
\item An experimental validation of proposed method is performed where simulated GPS time-stamp is shifted to model the timing attack. It has been found that timing attack can change the low-rank approximation profile of Henkel matrix created with unwrapped phase angle data.
\end{itemize}

This paper is organized as follows: section II discusses the low rank approximation technique of Henkel matrix. section II discusses the use of unwrapped phase angle data in distinguishing timing attack from FDIA. Section IV describes the proposed timing attack detection algorithm. The experimental validation with IEEE 13 bus system is explained in section V.

\section{Low Rank Henkel Matrix Structure}

Low rank approximation of synchrophaoor data has been used in various applications such as recovering missing data \cite{4}, event identification \cite{3} and cyber-attak detection \cite{5}. For a PMU data-set with m channel and n measurements, the matrix Y containing PMU measurement data can be expressed as a $m\times n$ matrix as follows:

$Y = \begin{bmatrix}
y_{11} & y_{12} & ... & y_{1n}\\
y_{21} & y_{22} & ... & y_{2n}\\
... & ... & ... & ...\\
... & ... & ... & ...\\
y_{m1} & y_{m2} & ... & y_{mn}
\end{bmatrix}$

If the Singular Value Decomposition (SVD) of Y can be written as $Y = U \Sigma V^*$, then it can be approximated as a rank r matrix ( r < rank(Y) ). This is done by taking the first r largest singular values in $\Sigma$, which is another diagonal matrix $\Sigma^r$. The low rank approximation error is:

\begin{equation}\label{e1}
    e^r = \frac{|| U \Sigma^r V^* - Y||_F}{  ||Y||_F} \times 100 \% 
\end{equation}

Ref \cite{5} exploits the low rank approximation of measurement vector Z received from PMUs to detect the unobservable False Data Injection attack. However, this method fails to address the time-series variation of PMU data under cyber attack. In order to address the temporal variation of data under attack, the low rank approximation of Henkel matrix can be utilized.\\
A Hankel matrix is defined as a square matrix which is constructed in such a way that each ascending skew-diagonal from left to right remains constant. Elements in each row are right-shifted data-streams from the previous row. For the datastream $a_0, a_1,.....a_k$, the constructed Henkel matrix will be:

$H = \begin{bmatrix}
a_0 & a_1 & ... & a_{k/2 +1}\\
a_1 & a_2 & ... & a_{k/2 +2}\\
... & ... & ... & ...\\
... & ... & ... & ...\\
a_{k/2 +1} & a_{k/2 +2} & ... & a_k
\end{bmatrix}$
\\
\\Henkel matrix H is a $(k/2 +2)$ $\times$ $(k/2 +2)$ square matrix. Henkel matrix has been proved to be useful for analyzing time-series data and state-space representations. The SVD of Henkel matrix can be utilized for decomposing the signal into temporal and spatial variation of signal \cite{1}. Low rank approximation of Henkel matrix can give insights to the attack on the time-series PMU data.\\
During an electrical event, there is a positive correlation of changes in phasor measurements between the neighboring PMUs \cite{6}. Therefore, a temporal relation exists between the channels during electrical events. When a random column permutation is performed, the temporal relation will be changed. As a result, the rank of the Henkel matrix will be higher.
During a False Data Injection Attack, only the measurements of affected PMU will change, the neighboring PMUs dont have any correlation with the affected PMU and there is no temporal relation between PMUs.
A random column permutation wont change the rank of the Henkel matrix in case of FDIA \cite{2}. The step by step process is explained in Algorithm 1.\\
Algorithm 1 is useful to detect the FDIA and to differentiate it from electrical events. Nevertheless, differentiating the FDIA and timing attack is yet to be explored. During timing attack, the phase angle values are modified and shifted. The FDIA can also modify the phase angles by adding or subtracting a specific amount from the phase angle measurements. As a result, the conventional attack detection schemes, which rely on the statistical deviation between the observed and actual measurements, fail to distinguish the phase angle measurements after FDIA and after timing attacks. One possible solution can be observing the unwrapped phase angle data instead of raw phase angle data.

\section{Unwrapped Phase Angle Data for Timing Attack Detection}

\begin{figure}
     \centering
     \begin{subfigure}[hbt!]{0.5\textwidth}
         \centering
         \includegraphics[width=\textwidth]{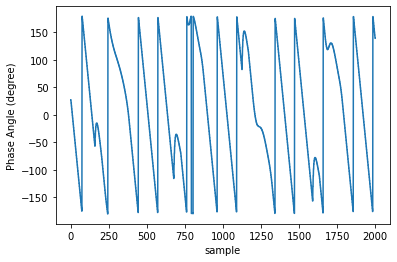}
         \caption{}
         \label{a1}
     \end{subfigure}
     \hfill
     \begin{subfigure}[hbt!]{0.5\textwidth}
         \centering
         \includegraphics[width=\textwidth]{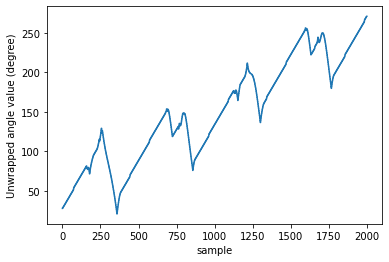}
         \caption{}
         \label{a2}
     \end{subfigure}
     \caption{Raw (a) and unwrapped (b) phase angle data from a random PMU }
\end{figure}

The PMU phase angle data deviate largely due to the fluctuations of synchronized system frequency around 60 Hz. To meet the IEEE C37.118 synchrophasor standard \cite{7}, the phase angle must be between $+\pi$ to $-\pi$. Thus the phase angle data wrapped around for an amount of $2\pi$ radian whenever it changes from $+/-\pi$ to $-/+\pi$. To resolve this issue, angle unwrapping techniques have been developed. An efficient real-time angle unwrapping technique was implemented in \cite{9}, which suffers the problem of making the unwrapped phase angle grow large over time. \cite{8}. To avoid the problem of unwrapped angle growing too large, a Roll-Over Counter (ROC) based unwrapping algorithm was proposed in \cite{8}. This algorithm keeps track of the number of times the phase angle shifts from $+/-\pi$ to $-/+\pi$, which is defined as ROC. ROC, in turn, is used to calculate the unwrapped angle by adding or subtracting an offset value which minimizes the difference between two consecutive phase angle data.\\
If two consecutive phase angle data are $\theta_i$ and $\theta_{i+1}$, then the ROC will be added by a an integer N. The N is defined to be the following minimizer:

\begin{equation}\label{e2}
\begin{aligned}
& \underset{N}{\text{min  }}
|\theta_{i+1} - \theta_i + 360N|  \\
\end{aligned}
\end{equation}
\begin{equation}\label{eroc}
    ROC(i+1) = ROC(i) + N\\
\end{equation}
N is 1 when the phase angle face transition from $+\pi$ to $-\pi$ and N is -1 when phase angle face transition from $-\pi$ to $+\pi$. The ROC(i+1) is calculated by adding the N with the previous ROC value (eqn \ref{eroc}). Fig\ref{a1} represents raw phase angle data from a random PMU and Fig \ref{a2} represents corresponding unwrapped phase angle data.\\
As mentioned before, unwrapped phase angle data is useful to avoid the wrapping-up phase angle during the transition of 360 $^o$. Our goal is to observe the behavior of phase angle unwrapping technique described in \cite{8} during cyber attack on a PMU. For the case of False Data Injection Attack (FDIA), the attacker modifies the measurement value directly. For the time instance t, the attacker change the phase angle in eqn \ref{e2}, changing $\theta(t)$ to $\theta'(t)$/ by adding an attack value a(t). The attacker makes the similar change for the following time instance t+1 by adding attack value a(t+1). Assuming original $\theta(t)$ is $\sim +180$ and $\theta(t+1)$ is $\sim -180$. The new phase angle values will be as follows:

\begin{equation}\label{e3}
    \begin{gathered}
    \theta'(t) = \theta(t) + a(t)\\
    \theta'(t+1) = \theta(t+1) + a(t+1)
    \end{gathered}
\end{equation}

From eqn \ref{e3} and fig \ref{a1}, we can observe that adding an attack vector to the phase angle data will increase or decrease the angle value, but will not impact the instance when there is a transition between positive half-cycle to a negative-half cycle. Our assumption is that the attack will try to modify the phase angle with a value which is not big enough to be easily detected by the defender. As a result, the attacker wont change the transition point between +180 to -180. From eqn \ref{e2}, it is evident that the N value remains same as the transition status between positive and negative is not changed. Thus the ROC value will be same, which makes the unwrapped angle graph during FDIA similar to the unwrapped angle graph during normal condition.
\begin{algorithm}[t!]
\SetAlgoLined
 \textbf{Initialization}\: Receive time-series measurements from PMU as the matrix Y. m is the number of channel, n is the data length of each PMU;\\
 Step 1: Create a $m(n/2 +2)$ $\times$ $(n/2 +2)$ Henkel matrix H;\\
 Step 2: Calculate the low rank approximation error $e^r$ with varying rank r ($r \leq rank(H)$);\\
 Step 3: Do a random column permutation on the Henkel matrix H and create a new matrix $\bar{H}$;\\
 Step 4: Calculate the low rank approximation error $e^{rr}$ with varying rank r ($r \leq rank(\bar{H}))$;\\
 Step 5: If $e^{rr} > e^r$, it is an electrical event; \\
 Step 6: else, it is a FDIA;\\
 \caption{Distinguishing cyber attack from electrical event in PMU}
\end{algorithm}

On the other hand, when there is a timing attack, the phase angle value will be shifted toward the horizontal time axis. Therefore the transition point between +180 to -180 will no longer be at the same point. If the time-shifted phase angle value is $\theta''$ and the original phase angle value is $\theta$, we can express the relation between  $\theta''$ and $\theta$ as follows:
\begin{equation}\label{e4}
    \theta''(t) = \theta(t+T)\\
\end{equation}

T is the amount of time-delay occurred due to timing attack. The data the Control Center receive at the $t^{th}$ sec is actually the data that the power grid generated T seconds ago. Since the transition point between +180 and -180 is changed, the N value from eqn \ref{e2} will also change at the time instance t. Different N will give different ROC value from eqn \ref{eroc}. As a result the unwrapped angle curve from fig \ref{a2} will not be able to maintain the similar shape and will be distorted.\\
From the above discussion, it can be concluded that we can exploit the behavior of the unwrapped phase angle curve to distinguish between the FDIA and timing attack. If the curve shape remain unchanged, even though the values of phase angle might change, it implies FDIA. On the other hand, if the unwrapped phase angle curve is distorted, we can conclude that it is timing or GPS spoofing attack.

\section{Using Low Rank Henkel Matrix Property to Detect Timing Attack}

According to the discussion in section III, distortion in the phase angle curve indicates timing attack. However, this criterion alone is not enough for timing attack detection. Distortion in phase angle curve may occur even in normal condition. Such as the change in functionality of the BUS that the corresponding PMU is connected to can lead to a variation in unwrapped angle curve shape. Moreover, any electrical event can change the phase angle data randomly, which may cause distortion in unwrapped angle curve. So there is a need to specifically identify the timing attack occurrence.\\
Random column permutation in low rank Henkel matrix can distinguish between event and cyber attack. As mentioned in section II, if the rank of Henkel matrix is different than its low rank approximation after random column permutation, then it is an event, and if the rank doesnt change after random column permutation, then it is an attack.

\begin{table}[hbt!]
\centering
\caption{Timing Attack Detection Criterion}
\label{t1}
\begin{tabular}{ | m{4em} | m{2cm}| m{2cm} |m{2cm} | } 
  \hline
  Incident& Random column permutation in Henkel matrix & Unwrapped phase angle curve distortion & Combination of both \\ 
  \hline
  Event & Rank changes & Curve shape distorted & Higher rank after column permutation on the distorted unwrapped phase angle curve \\ 
  \hline
  FDIA & Rank doesnt change & Curve undistorted & Same rank after column permutation on the undistorted unwrapped phase angle Henkel Matrix \\ 
  \hline
  Timing attack & Rank doesn't change & Curve distorted & Higher rank after column permutation on the distorted unwrapped phase angle Henkel Matrix \\ 
 \hline
\end{tabular}
\end{table}

From table \ref{t1}, it can be observed that exploiting both the random column permutation of Henkel matrix and the distortion in unwrapped phase angle curve can help distinguishing the timing attack from FDIA and electrical events. To identify the unwrap angle curve distortion, the low rank approximation of Henkel matrix can again be utilized. Since the FDIA doesn't change the shape of unwrap phase angle curve, the temporal relation between all the PMU data remain same. However, during timing attack, the affected PMU curve gets distorted. As a result, the temporal relation between the affected PMU and the other neighbouring PMUs doesn't remain same anymore. It leads to a higher value while performing low rank approximation. The proposed method is described in algorithm \ref{alg2}.

\begin{algorithm}[hbt!]

\SetAlgoLined
 \textbf{Initialization}\: Receive time-series angular (voltage or current) measurements from PMU as the matrix Y. m is the number of channel, n is the data length of each PMU;\\
 Step 1: Create a $m(n/2 +2)$ $\times$ $(n/2 +2)$ Henkel matrix H;\\
 Step 2: Calculate the low rank approximation error $e^r$ with varying rank r ($r \leq rank(H)$);\\
 Step 3: Do a random column permutation on the Henkel matrix H and create a new matrix $\bar{H}$;\\
 Step 4: Calculate the low rank approximation error $e^{rr}$ with varying rank r ($r \leq rank(\bar{H}))$;\\
 Step 5: If $e^{rr} > e^r$, it is an electrical event. Go to step 1 with the next n data sample; \\
 Step 6: else, it is an attack;\\
 Step 7: Unwrap the angular data and create Henkel matrix $H_u$ similar to step 1;\\
 Step 8: Calculate the low rank approximation error $e^{ru}$ with varying rank r ($r \leq rank(H_u)$);\\
 Step 9: if the low rank approximation error $e^{ru} > e^r$, then it is a timing attack;\\
 Step 10: Else, it is FDIA\\
 \caption{Detection of Timing Attack}
 \label{alg2}
\end{algorithm}

\section{Experimental Results}

\begin{figure}
  \includegraphics[width= 0.5\textwidth]{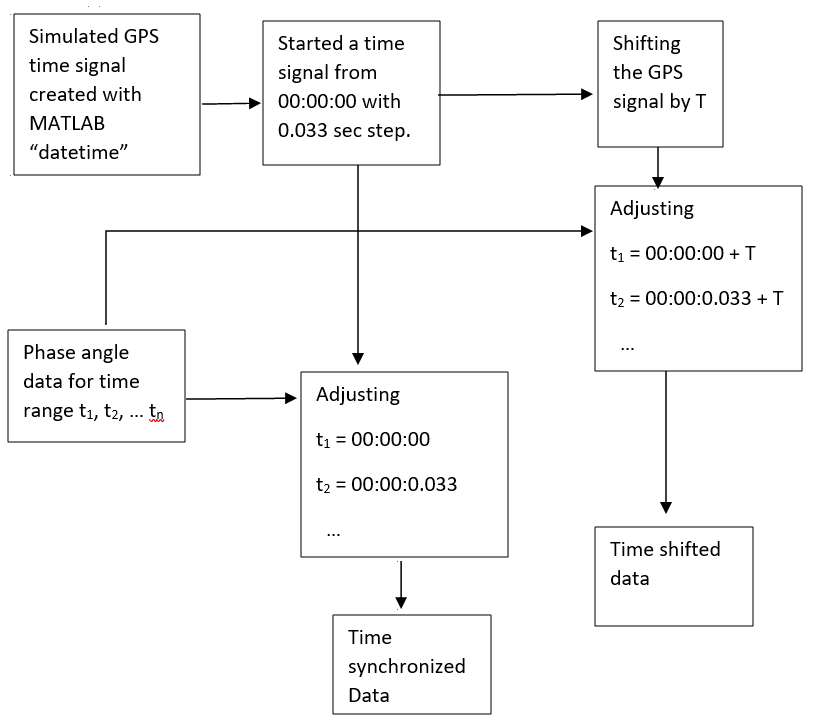}
 \caption{Block diagram of the simulation setup to analyze timing attack}
\label{b_1}
\end{figure}

The aforementioned timing attack detection method is tested in the IEEE 13 bus system. The data used in this work spans over 1 hour period. To simulate the timing attack, we have created an experimental setup, where the GPS timestamp is simulated using MATLAB \textit{datetime} function. It provides a series of universal time reference (UTC) timestamp beginning from a specified point of date and time. In this work the time-stamp midnight (00:00:00) with a sampling rate of 1/30 sec. The first data is at 00:00:00, the second data is at 00:00:0.033, and so on. The last data of a single day is at 23:59:59.\\
\begin{figure}
  \includegraphics[width= 0.5\textwidth]{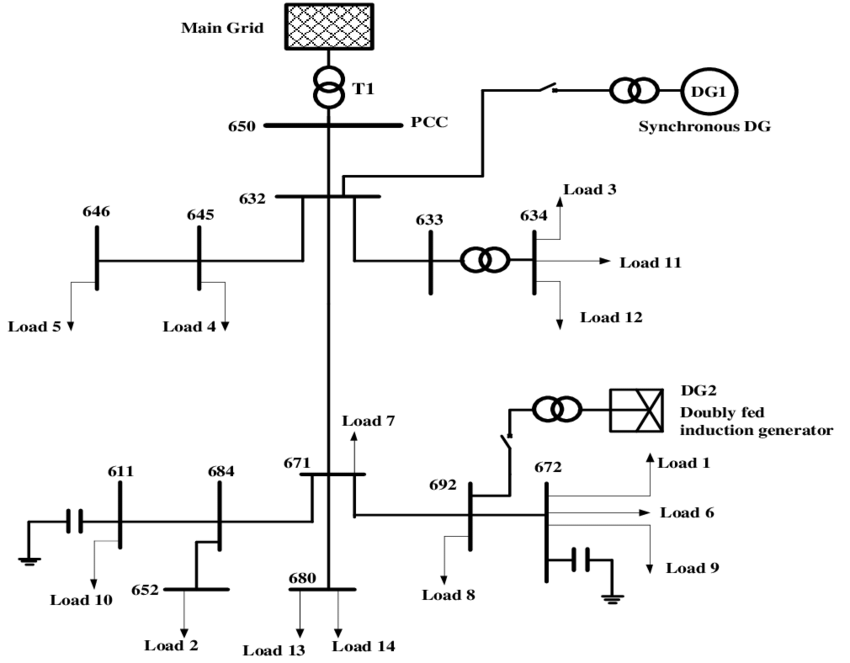}
 \caption{IEEE 13 Bus system}
\label{13}
\end{figure}
The simulation of IEEE 13 Bus system has been done with MATLAB SIMULINK. The PMUs have been added at buses 632, 633, 634, 671, 672 and 692 (fig \ref{13}). The PMUs provide positive sequence voltage magnitude and angle, positive sequence current magnitude and angle and frequency data. We have considered the positive sequence voltage angle data and synchronized the data with simulated GPS time-stamp mentioned before. After this step, we have shifted the GPS time-stamp by a time T, so that the voltage angle values are synchronized with incorrect and shifted time-stamp. In this way the timing attack on PMU can be simulated. The block diagram of proposed time-stamp generation and creating a timing attack is illustrated in fig \ref{b_1}. \\
A time shift of T = 3 sec is applied onto the GPS timestamp to analyze timing attack. To demonstrate the FDIA, a random attack vector \textbf{a} ranging from 0 to 30 has been added to the PMU data-stream. In each case, the unwrapped angle has been calculated and the corresponding behavior with respect to time is observed.\\
\begin{figure}
  \includegraphics[width= 0.5\textwidth]{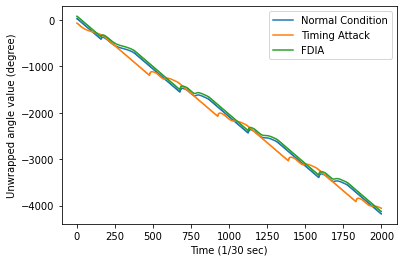}
 \caption{Unwrapped phase angle data during normal condition, under timing attack and under False Data Injection Attack}
\label{b_2}
\end{figure}
From fig \ref{b_2} it is evident that during timing attack the phase angle curve is distorted, whereas during FDIA the unwrap angle curve maintains similar shape at the moment of attack despite being shifted upward. From section IV, the distortion alone cannot indicate if there is any kind of performance issue in the grid i.e. event, FDIA or timing attack. Therefore according to the step 7 of algorithm \ref{alg2}, a Henkel matrix is created with PMU data. Here the total number of channels is m = 6 and the data-length is n = 100. The number of rows in Henkel matrix is $6 \times (100/2 +2) = 312$. The low rank approximation error profile has been created as in eqn \ref{e1}.\\
\begin{figure}
  \includegraphics[width= 0.49\textwidth]{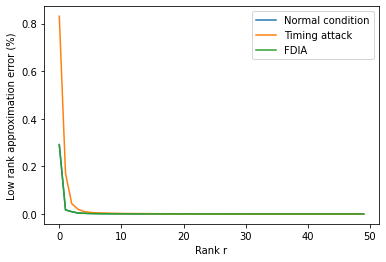}
 \caption{Low rank approximation error under normal condition, under FDIA and under timing attack}
 \label{r1}
 \end{figure}
As the temporal relation between PMU channels get disrupted after timing attack, the low rank approximation error should be higher for timing attack, though the FDIA should exhibit no change in the low rank approximation error profile. Fig \ref{r1} confirms this theory. Here the low rank approximation error is higher for timing attack, where there is no visible change during the FDIA.\\
\begin{figure}
  \includegraphics[width= 0.49\textwidth]{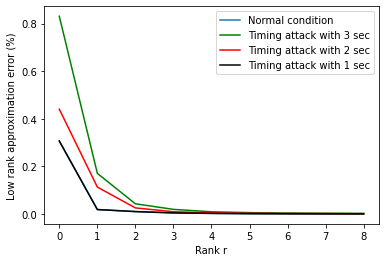}
 \caption{Low rank approximation error under normal condition, under FDIA and under timing attack}
 \label{r2}
 \end{figure}
For a 3 sec shift in the time-stamp reference, the low rank approximation error increases by 700 $\%$ (fig \ref{r1} for r = 1). This increase gets reduced as the rank r increases. After r = 5, the change in low rank approximation error is insignificant. The increase in low rank approximation error is smaller for a 2 sec shift in the time-stamp reference. One sec time shifting doesn't create significant impact even at r = 1. Therefore, it can be concluded that different timing attack will result in different low rank approximation error of Henkel matrix.

\section{conclusion}

Introduction of PMUs to cyber-physical system provides accurate data acquisition, while posing additional risk of being the victim of cyber attack. Both False Data Injection Attack (FDIA) and GPS-spoofing or timing attack can provide malicious data to the cyber system, though these two attacks require different post-attack contingency plan. Thus accurate detection of timing attack and separating it from conventional FDIA has become a very important research area. In this article, a successful detection of timing attack mechanism is proposed. Firstly, a method to distinguish timing attack and FDIA using unwrapped phase angle data is developed. Secondly, utilizing low rank Henkel matrix property to differentiate timing attack from electrical events is also presented. Finally, an experimental validation of proposed model is performed on IEEE 13 bus system using simulated GPS-spoofing attack. It can be observed that the timing attack can increase the rank 1 approximation error of Henkel matrix of unwrapped angles by 700 $\%$ for 3 sec delay in GPS time-stamp. The rank 1 approximation error is increased by 500$\%$ for 2 sec delay and the increase is insignificant for 1 sec delay attack. FDIA doesn't show any significant change in the low rank approximation profile of Henkel matrix.

\bibliography{references} 
\bibliographystyle{ieeetr}

\end{document}